# Calcium carbonate polyamorphism and its role in biomineralisation: How many ACCs are there?

*Julyan H. E. Cartwright[1]\*, Antonio G. Checa[1,2]\*, Julian D. Gale[3]\*, Denis Gebauer[4]\*, C. Ignacio Sainz-Díaz[1]\**

1. Instituto Andaluz de Ciencias de la Tierra, CSIC–Universidad de Granada, E-18100 Armilla, Granada, Spain; julyan.cartwright@csic.es (JHEC); sainzdiaz@gmail.com (CISD)

2. Departmento de Estratigrafía y Paleontología, Universidad de Granada, E-18071 Granada, Spain; acheca@ugr.es

3. Nanochemistry Research Institute, Department of Chemistry, Curtin University, PO Box U1987, Perth, WA 6845, Australia; J.Gale@curtin.edu.au

4. Department of Chemistry, Physical Chemistry, University of Konstanz, D-78457 Konstanz, Germany; denis.gebauer@uni-konstanz.de

ABSTRACT

While the polymorphism of calcium carbonate is well known, and its polymorphs — calcite, aragonite, and vaterite — have been much studied in the context of biomineralisation, polyamorphism is a much more recently discovered phenomenon, and the existence of more than one amorphous phase of calcium carbonate in biominerals has only very recently been understood. Here we summarise on the one hand what is known of polyamorphism in calcium carbonate, and on the other what is understood of amorphous calcium carbonate in biominerals. We show that considering the amorphous forms of calcium carbonate within the physical notion of polyamorphism leads to new insights when it comes to the mechanisms by which polymorphic structures can evolve in the first place. This has not only implications for our understanding of biomineralisation, but also of the means by which crystallisation may be controlled in medical, pharmaceutical or industrial contexts.

# 1. Introduction

Calcium carbonate is one of the materials that has been found to possess more than one amorphous phase; a phenomenon termed amorphous polymorphism, or polyamorphism[1]. While polymorphism — the possession of more than one crystalline phase — is a long-studied phenomenon, polyamorphism is a much more recently discovered occurrence. The majority of research on polyamorphism to date relates to water, where three ice polyamorphs, termed LDA, HDA, and VHDA (low-density amorphous; high-density amorphous, and very-high-density amorphous) have been identified[2–4] (Figure 1). Apart from ice, another case of polyamorphism that has been studied is in silicon[5]. While the majority of systems that exhibit polyamorphism possess strongly directional interactions, such as covalent or hydrogen bonding, it has also been shown to occur in metallic alloys[6]. Furthermore, polyamorphism can also occur in the liquid state, the liquid being a dynamic version of an amorphous solid state; such liquid polyamorphism is seen in phosphorus[7] and triphenyl phosphite[8].

Here we discuss the occurrence of polyamorphism in calcium carbonate, a mineral that is well known both for its polymorphism, and also for its importance in biomineralisation; that is, in mineral structures formed by biological systems. We believe that it is timely to examine polyamorphism in calcium carbonate from the point of view of biominerals, since on the one hand amorphous calcium carbonate (ACC) is increasingly acknowledged to be playing an important role in biomineralisation (Section 2), and on the other, it is becoming increasingly clear that there is not just one form of ACC, but that there exist polyamorphs (Section 3). In addition, new insights into both ACC and the highly amorphous precursor species have recently come to light from molecular dynamics simulations (section 4). We also highlight several open problems and challenges to be taken up (Section 5).

## 2. Amorphous calcium carbonate in biomineralisation

The notion that many organisms produce amorphous minerals such as silica, calcium phosphate, or calcium carbonate (ACC) has a long history, but it was not until the late 1960s that the first transient amorphous phase was identified in chiton teeth[9]. Since then there has been increasing evidence that biominerals, both in vertebrates and invertebrates, form via the corresponding amorphous precursor (see reviews in refs.[10–12]). Calcium carbonate is, by far, the most commonly employed material by invertebrates for the construction of hard structures. These occur as granules, spicules, shells and additional structures, and can be made of either of the two main calcium carbonate biominerals, aragonite or calcite as well as, more rarely, vaterite.

According to Addadi et al.[13] the first mention of biogenic ACC was made in the early 20th century, although the first detailed studies involving high resolution techniques began in the 1990's. Since then, ACC has been widely recognized in many groups of organisms (see review in Addadi et al.[13]). In some of them, ACC is used as a structural component (e.g. plant cystholiths[14], calcitic sponge spicules[15–17], ascidian spicules[16–18]) or as a reservoir for future calcium carbonate availability (earthworms[19–21], arthropods[22–26]). In other groups (e.g., molluscs[27–33], sea urchins[34–42]), ACC is used as a precursor for the formation of crystalline phases of $CaCO_3$ (crystalline calcium carbonate; CCC). The study of the two latter groups is particularly intense due to the intrinsic general interest in processes involving the crystallization of ACC.

A matter of debate is how such a metastable phase as ACC becomes stabilised, either transitorily or permanently. There have been proposals that macromolecules, water, membranes and ionic components could perform this role[43]. Aizenberg et al.[15,16,18] found significant differences in the amino acid impurities of ACC and calcite from ascidian and calcitic sponge spicules, and proposed that macromolecules, in cooperation with $Mg^{2+}$, were responsible for ACC stabilisation. This view was later taken by other authors to explain ACC stabilisation in crustaceans[23], sea urchin larval[38] and adult[35,42] spines, and earthworms[19,20]. More recently, the role of low molecular weight metabolites (inorganic phosphates, phosphoenolpyruvates, citrates…) has been demonstrated for the stabilisation of the ACC of crustaceans[26,44].

Beniash et al.[34] were the first to propose the transformation of ACC into another crystalline phase (high Mg calcite in the larval spicules of the sea urchin). This interpretation was later reinforced by the detection of granules of ACC in spiculogenic cells, which could subsequently be transported to the mineralisation site[35]. Similarly, Hasse et al.[27] and Marxen et al.[29] in the gastropod *Biomphalaria*, Weiss et al.[28] in larval bivalves (although Kudo et al.[45] did not find ACC in the larval shell of a *Crassostrea*), and Jacob et al.[32] in freshwater cultured pearls, have implied transformation of ACC into aragonite. All the above authors observed how during growth of structures the ratio of CCC to ACC increased. This, *per se* does not imply transformation of ACC into CCC, because direct deposition of CCC might occur in more advanced growth stages.

The first direct evidence was obtained at the same time in two distant groups of invertebrates. Politi et al.[37] showed by differential etching how the ACC of the regenerating sea urchin spine occurs in a 100-200 nm thick outer layer (Figure 2A) and deposits preferentially at the tip, which is where the growth rate is the fastest (see also Seto et al.[42]). They also imaged the transformation of ACC into calcite under transmission electron microscope (TEM) irradiation. Simultaneously Nassif et al.[30] found a 3–5 nm ACC layer coating the surface of mature nacre tablets. A similar, though thicker and more continuous, amorphous front was found in the prismatic calcitic layer of the pearl oyster[31] (Figure 2B). Since all these structures grow by the addition of material at the growth front, the only possibility is that the ACC cortex transforms into the corresponding crystalline phase. Transformation of ACC into CCC immediately raises the interesting issue of how this transition takes place. During the study of the formation of the spicules of the sea urchin embryo, Beniash et al.[35] found that ACC contains significant amounts of water that are released during the crystallisation process.

Later work on this structure and on sea urchin spines[37,38,40] and teeth[41] made it clear that stabilised biogenic ACCs typically contain structural water, while transient biogenic ACCs were in general anhydrous (cf. section 4). Biomineralisation proceeds in three stages: a) formation of hydrated ACC, which rapidly transforms into b) short lived anhydrous ACC, which changes into c) calcite (see review in Addadi et al.[13]). According to experimental data[46], this sequence is energetically downhill.

Nudelmann et al.[47] first suggested that calcitic prisms of a bivalve grow by precipitation of ACC particles (50-100 nm diameter), which subsequently crystallise epitaxially upon contact with the crystalline surface. Politi et al.[40] were the first to propose that short-lived anhydrous ACC of sea urchin larvae transforms into calcite by secondary nucleation, in which crystallisation of ACC stimulates the transformation of the domains in contact. In this process, the two phases (ordered and disordered) are both solid and in contact with each other, and the transformation involves solid state transformation[48]. The crystallographic orientation is determined by that of the central initial crystal of the larval skeleton. The secondary nucleation hypothesis is supported by experiments, in which, using Langmuir monolayers as model systems, solid state transformation of ACC into vaterite has been observed[49].

Distinct short-range order has been observed in ACC. The first evidence came from ACC of the aragonitic freshwater snail *Biomphalaria glabrata*, which was found to have a short-range order similar to aragonite[27,29]. Other orders relating to aragonite[13] and calcite[13,38,43] were found in aragonitic larval molluscs and in calcitic sea urchin structures, respectively. Monohydrocalcite-like ACC has also been referred to in a series of other organisms[50]. Therefore the statement of Addadi et al.[13] that biogenic ACC is structurally not one mineral phase, but a family of phases that appear to be genetically controlled in different species and phyla, makes sense. Belcher et al.[51] and Falini et al.[52] concurrently demonstrated that polymorph secretion is controlled by the macromolecules associated with either calcitic or aragonitic shell layers. The same macromolecules could also control the short-range order of ACC, if we consider the polyamorph-polymorph relationship commented on above.

# 3. Proto-crystalline versions, or polyamorphs, of calcium carbonate

Apart from biogenic specimens, synthetic amorphous calcium carbonates (ACC) can exhibit distinct short-range structures, too. In the presence of poly(aspartic acid) and magnesium ions, short-range structures resembling vaterite and aragonite may be obtained, respectively[53]. Additives can stabilize ACC, which is unstable at ambient conditions, but additive-free stabilized ACC can be isolated from aqueous environments if it is precipitated from high levels of supersaturation[54–57]. However, such ACCs show no distinct short-range order[58,59]. These observations may imply that additives induce short-range structuring in ACC, i.e., that distinct local order would be an extrinsic feature of ACC. This notion is in accord with the observation of structuring in biogenic ACC, since this always contains bio(macro)molecules as well, and could hence be under genetic control as speculated by Addadi et al.[13]. However, it is now evident that distinct short-range structural features are intrinsic to ACC if it is precipitated from a moderate level of supersaturation.

Additive-free ACC that is precipitated from equilibrated, slightly supersaturated (metastable) solutions of calcium carbonate by means of a sudden change to a medium that only weakly solvates calcium carbonate (e.g. a "quench" in ethanol) exhibits distinct short-range order[60]. The structuring depends on the pH level, that is at pH ~8.75 and pH ~9.80 ACCs with short-range structures that relate to calcite and vaterite, respectively, are obtained (Figure 3). Different structures in ACC at their respective pH levels were suggested in an earlier study, as two different solubilities of ACCs were identified (ACCI and ACCII[61]). Solubilities directly correlate with the thermodynamic stability of the respective phases (and can reflect size, because Gibbs-Thomson effects become important at the nanoscale, or the presence of impurities). The more stable ACC (ACCI) resembles the short-range structure of calcite, which is the more stable crystalline polymorph, while the less stable ACC (ACCII), on the other hand, relates to the least stable anhydrous crystalline polymorph, vaterite[60].

Consequently, the notion of proto-crystalline structuring in ACC has been introduced; namely ACCI and ACCII are proto-calcite ACC (pc-ACC) and proto-vaterite ACC (pv-ACC), respectively[60].

Interestingly, the different stabilities of ACC reflect different, pH-dependant stabilities of pre-nucleation clusters[61]. The more stable pre-nucleation clusters correspondingly yield the more stable pc-ACC upon nucleation. Since nucleation appears to proceed via aggregation of the pre-nucleation clusters[61–63], this strongly suggests that distinct structures are present in pre-nucleation clusters, too[64]. This is evidence that the different structures in ACC are intrinsic and can depend on intensive parameters during the early stages of crystallisation. Moreover, this mechanism may rationalise why no distinct structures may be obtained in ACC if it is precipitated from very high levels of supersaturation (cf., above); in this case, ACC is precipitated virtually instantaneously, and proto-crystalline structuring in pre-nucleation clusters (and with it, in ACC) cannot equilibrate according to a given set of intensive variables. On the other hand, interactions with additives may stabilise certain proto-crystalline structures in ACC, which are successively developed. Recent modelling results indicate that chain-like and highly dynamic structures in pre-nucleation clusters may be the basic principle behind proto-crystalline structuring in intermediate ACC (see Section 4)[65].

Despite the above, the proto-crystalline structures of additive-free amorphous intermediates do not predetermine the outcome of the amorphous to crystalline phase transition[60]; proto-calcite ACC does not necessarily transform into calcite. Since additive containing and biogenic, proto-structured ACCs can transform in an unambiguous manner, it appears that additives may specifically interact with the proto-structures during the amorphous to crystalline phase transitions, and in this way control polymorph selection[66]. However, there are also exceptions to this rule; for instance, the 'aragonitic' ACC in the freshwater snail *Biomphalaria glabrata* can apparently transform into minor amounts of vaterite in adult animals[27]. The detailed mechanisms underlying additive-mineral interactions throughout the different stages of crystallisation (pre-nucleation, nucleation, post-nucleation) remain as yet unknown, however (see Section 5).

## 4. Simulation of the amorphous phases and "proto-crystalline" structures

Determining structural models for amorphous calcium carbonate is a challenging prospect experimentally. Recently it has been possible to arrive at two different sample atomic configurations based on Reverse Monte Carlo fitting to Pair Distribution Function data for ACC[56]. It should be noted that here ACC was precipitated from very high levels of supersaturation and consequently did not exhibit distinct proto-structural features (cf. section 3). While this is suggestive of a possible heterogeneous water distribution within ACC when the stoichiometry is close to that of monohydrocalcite, there remain uncertainties as to how representative these configurations may be.

An alternative approach that complements experimental studies is to exploit atomistic simulation techniques and, in particular, molecular dynamics to provide structural insights into ACC. Quigley and Rodger[67] have used metadynamics, a form of bias acceleration, to estimate the relative free energies of amorphous and crystalline nanoparticles of calcium carbonate. Although the underlying force field fails to capture the correct relative stability of the crystalline polymorphs, this work demonstrates that it is possible to overcome the limited timescales accessible to molecular dynamics that would normally prevent phase transformations from being observed.

Raiteri and Gale[68] have taken an alternative approach to the study of amorphous nanoparticles by quenching clusters that have been melted *in vacuo* and subsequently annealing them in an aqueous environment. Here a range of cluster dimensions have been probed, from ion pairs through to diameters approaching 4 nm, while also varying the water content. Although no clear evidence for polyamorphism was observed, there are pointers toward size-dependent structural inhomogeneity. In particular, the thermodynamically favoured water content was found to increase as particles grow. Combined with the hindered diffusion of water within ACC, it is probable that this will lead to radial variations in composition with the outer shell being wetter than the inner core. Furthermore, by optimising the water content, the free energy of ACC nanoparticles can remain lower than that of crystalline calcite nanoparticles at small sizes leading to ACC being stable, rather than metastable. However, as particles agglomerate or grow they will rapidly become metastable with respect to calcite and aragonite.

Aside from the structure of ACC, molecular dynamics simulations can also play a role in understanding the nature of pre-nucleation species, as identified experimentally. Recently Demichelis et al.[65] have shown that calcium and carbonate ions in solution rapidly aggregate to form stable clusters. These precursors have an unusual and very dynamic structure consisting of chains of alternating cations and anions. Remarkably the system can adopt configurations in which the ions form rings, branched and linear chains that all possess the same free energy to within ambient thermal energy. This new type of species has been labelled a Dynamically Ordered Liquid-Like Oxyanion Polymer (DOLLOP), and has been suggested to represent the structural form of pre-nucleation clusters[65]. Although the behaviour of DOLLOP is not formally polyamorphism, as there is no phase boundary, these structures represent disordered clusters, but with different topologies that can interconvert (Figure 4). If these species, which are in equilibrium with solution, can reach a critical size it appears that they undergo a subtle change in structure that causes larger clusters to appear that are more stable than the initial DOLLOP form. However, it remains unclear what exactly happens at the point of nucleation, that is, how the DOLLOPs grow to reach a critical size; this process could in principle be based on either ion-by-ion growth or on aggregation of individual DOLLOPs. Experimental observations[61–63] point towards the latter pathway. Thus, the structural change in larger DOLLOPs, which can exhibit different interconverting topologies at smaller sizes, may be related to the proto-crystalline structures observed experimentally, thereby laying the foundation for polyamorphic ACC.

## 5. Open questions and challenges

There are many gaps in our knowledge of the polyamorphism of calcium carbonates, especially when it comes to understanding what is occurring at the atomic scale. The presence of strong electrostatic interactions between carbonate anions and calcium cations, and the hydrogen bonds owing to the presence of water molecules, must be the key to understanding the phase transitions between glassy forms, and also between amorphous and crystalline forms. Similar cases have been studied, such as in the polyamorphism of ice[69], and in the transformation of carbonate to carbonic acid[70] where several polymorphs have been detected. It is often assumed that each amorphous form yields one specific crystalline form of carbonate, however, this is not necessarily the case, but instead depends on the nucleation and crystallisation conditions. One of the main challenges in calcium carbonate polyamorphism is the lack of a structural model that describes these glassy phases. This would allow us to study the atomic arrangements of different crystallographic long-range ordered structures or different disordered structures, and the role of water and/or protein molecules in influencing the mechanism of the amorphous/crystal phase transition. The knowledge of a structural model for these amorphous phases would help in the understanding of the role of additives in biocrystallisation processes and the interactions occurring during the amorphous-crystalline phase transition. Various techniques used in the characterisation of crystalline forms of carbonates have also been used to distinguishing between the amorphous phases, such as infrared spectroscopy, X-ray diffraction, extended X-ray absorption fine structure (EXAFS), and synchrotron X-ray total scattering methods. However, only indirect information related to the atomic structure has been obtained so far[56].

EXAFS analyses of glassy calcium carbonates have found average coordination numbers for $Ca^{2+}$ cations lower than those existing in the crystal forms (6 for calcite and vaterite, and 9 for aragonite)[58]. Ca *K*-edge EXAFS experiments along with reverse Monte Carlo simulations have found different Ca distributions and water molecule environments within the amorphous carbonate structure according to nuclear magnetic resonance (NMR) and infrared spectroscopy studies[56]. However, no clear differences between the atomic structures of the possible amorphous phases have yet been detected. Indeed, it remains uncertain as to how homogeneous the composition, and therefore the structure, of partially hydrated ACC actually is.

Taking all of the above into account, we can conclude that the polyamorphism of calcium carbonate is rather extensive (Figure 5). To begin with, there are hydrous and anhydrous ACCs, which have been observed to correspond to stabilised and transient ACCs in biogenic specimens, respectively[13]. The water content in ACC can vary, while observations of stoichiometric, or near-stoichiometric $CaCO_3 \cdot H_2O$ preponderate[11]. Dehydration of water-containing ACCs toward anhydrous ACC, and subsequent crystallisation, follows a downhill energetic pathway[46], at least at larger particle sizes, which might indicate that transient biogenic ACCs are actually formed from stabilised hydrous precursors during biologically induced crystallisation. Besides the varying water content, biogenic ACCs can exhibit distinct short-range structural features, which can relate to calcite[13,38,43] (calcitic ACC), aragonite[13,27,29] (aragonitic ACC), or monohydrocalcite[50] (MHC-like ACC); vateritic biogenic ACC, however, has not yet been reported (Figure 5). Following our considerations above, the differently structured biogenic ACCs may exhibit varying states of hydration, depending on their biological function (storage or transient forms that are about to crystallise).

How the "pre-structured" biogenic ACCs relate to synthetic proto-structured ACCs (proto-calcite ACC, pc-ACC; proto-vaterite, pv-ACC[60]; and possibly proto-aragonite, pa-ACC, which has not been obtained yet (Figure 5)), in terms of polyamorphism is a matter of debate. On one hand, the origin of distinct short-range structures may also be based on pre-nucleation clusters and DOLLOP, as outlined in sections 3 and 4, in the biogenic case. On the other hand, structural characterisations by means of EXAFS show that biogenic species of ACC can have coordination numbers commensurate with crystalline species[71], as opposed to the low coordination of $N = 2$ in the synthetic proto-structured ACCs[60]. While comparably low coordination numbers can also be found in biogenic ACCs[50], this observation may in the end relate to the high uncertainty surrounding determining reliable coordination numbers by means of EXAFS[72]. An alternative explanation could be that bio(macro)molecules incorporated in biogenic phases of ACC could stabilise amorphous states that are already much closer to crystalline states, but which have originated from less ordered, though proto-structured, ACCs with lower coordination. This could also be true for distinctly structured ACCs obtained utilizing different additives *in vitro*[53].

Additive-containing ACCs that do not relate to crystalline polymorphs can be synthesised *in vitro*, notably the polymer-induced-liquid precursors (PILP)[11,73], and liquid ACC[74] that may be stabilised also in the absence of additives if the contact with an extrinsic surface, and thereby heterogeneous nucleation, is reduced through levitation in droplets[75]. Last but not least, additive-free ACC that does not resemble any crystalline polymorph can be obtained from high levels of supersaturation (labelled "unstructured" in Figure 5)[54–59], and stabilised without the help of additives; it may be speculated that indeed the high degree of structural disorder of this metastable phase leads to an intrinsic kinetic stabilisation against crystallisation.

Another open question related with ACC and carbonate crystallisation is to understand the effect of the presence of $Mg^{2+}$ cations on the formation of calcium carbonates and dolomites. Natural dolomite (calcium-magnesium carbonate) was formed at low temperature but it is not yet possible to synthesise it in the laboratory at low temperature. The main accepted explanation is that natural dolomite must have somehow originated from biomineralisation. To understand the formation of dolomite at low temperature, the prior formation of an amorphous carbonate phase should be taken into account[76]. This amorphous calcium (magnesium) carbonate can be stabilised with the presence of inorganic or organic additives that can also modulate the formation of different polyamorphs. The presence of $Mg^{2+}$ can favour environments with high water content or with low-coordination-number cations[56] stabilising some atomic arrangements that can be considered as amorphous phases. An understanding of the role of amorphous carbonate in dolomite formation could change previous interpretations of the origin of many dolomites[77]. It is known that the presence of $Mg^{2+}$ induces the formation of aragonite without incorporation of $Mg^{2+}$ in the crystal lattice. Probably the $Mg^{2+}$ cations favour the proto-aragonite ACC phase. We also should not forget the presence of organic molecules during the biomineralisation process; these organic additives can have carboxylate functional groups that can modulate the biogenic ACC to form a specific amorphous phase (see above). This organic matter with carboxylate groups can facilitate the ordering of the ACC atoms to nucleate a specific polymorph of crystalline carbonate. This effect can be combined also with that of $Mg^{2+}$ cations leading to a change in the amorphous precursor phase or the nucleation of crystalline form.

Perhaps it is not surprising that determining structural models for polyamorphic calcium carbonate is challenging when the details of the structure for even the crystalline (Figure 5, inset), but disordered, phase vaterite are still open to debate. Although it is widely accepted that vaterite has hexagonal symmetry on average, the arrangement of the carbonate groups within the structure is still a matter for discussion. While the carbonates are often believed to be disordered by rotation about the *c*-axis of the hexagonal cell, recent *ab initio* calculations[78] have indicated that most of the existing ordered structure models are dynamically unstable and that the carbonate groups also undergo subtle rotations about axes that lie in the *ab*-plane too.

Our hope is that this review will contribute to bringing together research being carried out into biogenic and synthetic ACC polyamorphism, so that a future review will be able to display a figure like Figure 1 for calcium carbonate.

# 6. Acknowledgments

We acknowledge funding provided by projects CGL2010-20748-CO2-01 and FIS2010-22322- C02-02 of the Spanish Ministerio de Ciencia e Innovación and the European COST Action TD0903. JDG thanks the Australian Research Council for a Professorial Fellowship and funding from the Discovery program. DG thanks Helmut Cölfen for his support.

# 7. Author Contributions

JHEC and DG developed the idea for this article, and all authors discussed the subject. All authors contributed to the writing and editing of the manuscript.

## 9. Figure Legends

**Figure 1** | Polyamorphism in ice: sketch of the polyamorphism "phase diagram" of water. Dotted lines: (upper) minimum crystallisation temperature of (super)cooled water and (lower) maximum crystallisation temperature of amorphous ices[79]. Filled blue circle: proposed second critical point of water and, below this, the (dot–dashed) proposed line of first-order transitions in (inaccessible) water[79,80], continuing into the estimated equilibrium phase boundary between LDA and HDA[81,82]. The arrows show the observed low- to high-density transition at 0.35 GPa, and reverse transition at 0.05 GPa back to LDA, at 130–140 K, crossing the upstroke and downstroke lines found by Mishima[81], shown as thin dashed lines. (On compression, LDA remains metastable up to the upstroke line, and, on decompression, HDA remains metastable down to the downstroke line.) Thick dashed line: approximate P–T boundary for formation of VHDA. Reproduced from ref.[83].

**Figure 2** | Evidence of ACC surrounding a regenerating sea urchin spine (A) and growing calcitic prism of a bivalve (B). A. Regenerating spine of the sea urchin *Paracentrotus lividus*. The lower left view is of a freshly formed microspine after 4 days of regeneration. The lower right view is of a similar microspine etched in water; dissolution of ACC has occurred in the external layer. Reproduced from ref.[37]. B. Growing front of a prism of the outer calcitic layer

of the bivalve *Pinctada margaritifera* showing the contact between the crystalline interior and the amorphous cortex. Reproduced from ref.[31].

**Figure 3** | Spectra of calcite, vaterite, proto-calcite ACC (pc-ACC), and proto-vaterite ACC (pv-ACC). a) $^{13}$C solid-state NMR spectra recorded by single pulses at a magnetic field of 9.4 T and a MAS rate of 8.0 kHz. b) Fourier Transform of calcium *K*-edge EXAFS plotted as a function of distance, *R*; the expected species assignments for the first three coordination shells are indicated. The black arrow marks a peak that may relate to the coordination of structural water. The vertical lines are a guide to the eye. The data show that pc-ACC and pv-ACC relate to the structure of calcite and vaterite, respectively, on average within the short-range structure. Reproduced with alterations from ref.[60].

**Figure 4** | Structures of the four formula unit pre-nucleation clusters. Structures shown represent the configurations of four separate clusters after 1 ns of simulation at experimental conditions ([Ca]=0.4 mM, [HCO$_3^-$]= 10 mM, pH=10). Atoms are coloured green, blue and red for calcium, carbon and oxygen, respectively. Reproduced from ref.[65].

**Figure 5** | Overview of calcium carbonate polyamorphism and polymorphism (inset). For explanations see text.

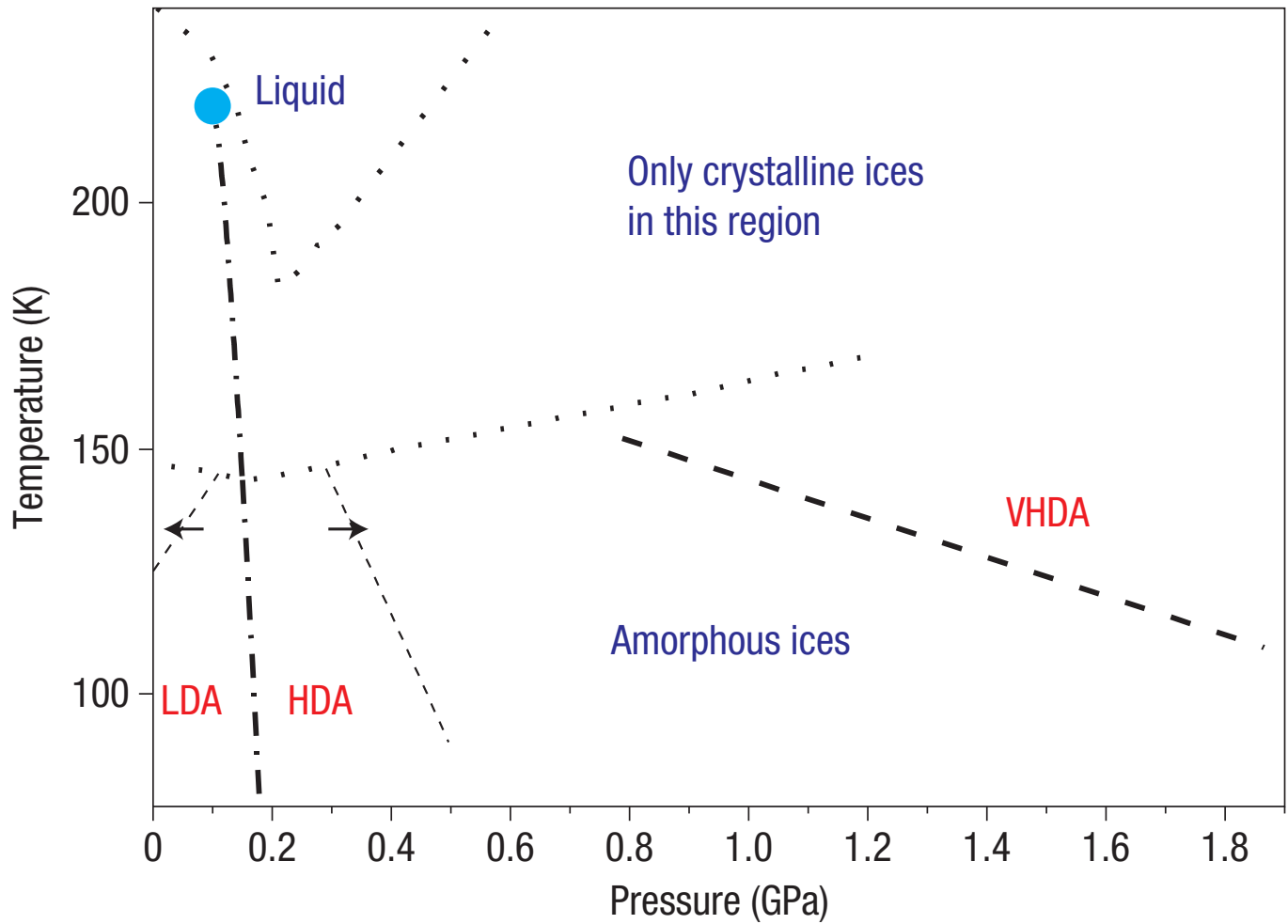

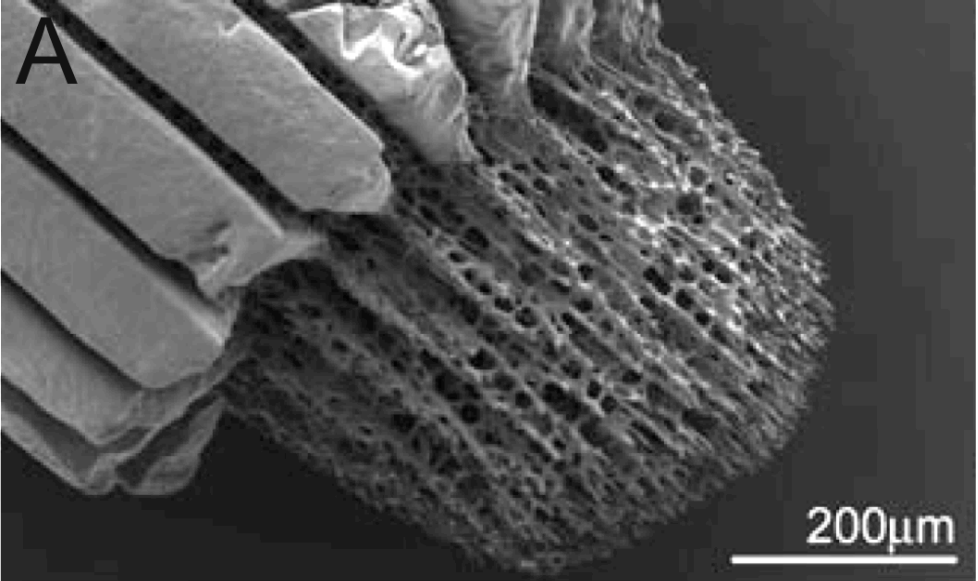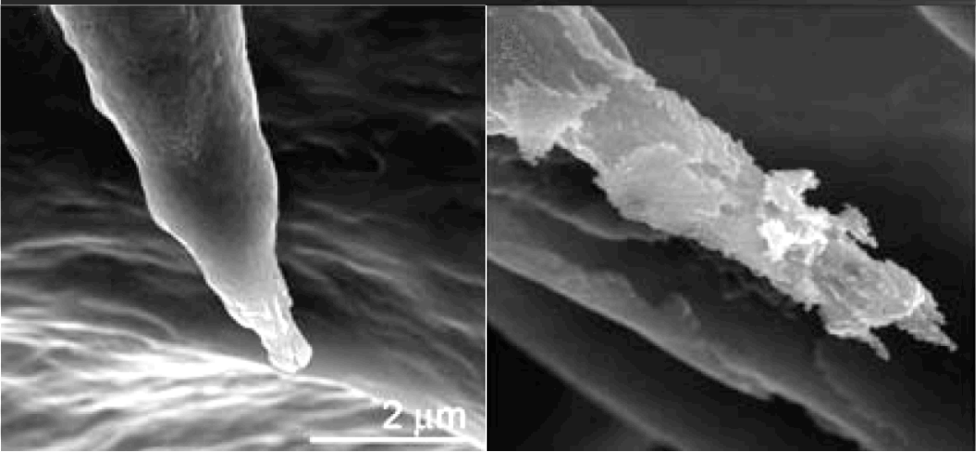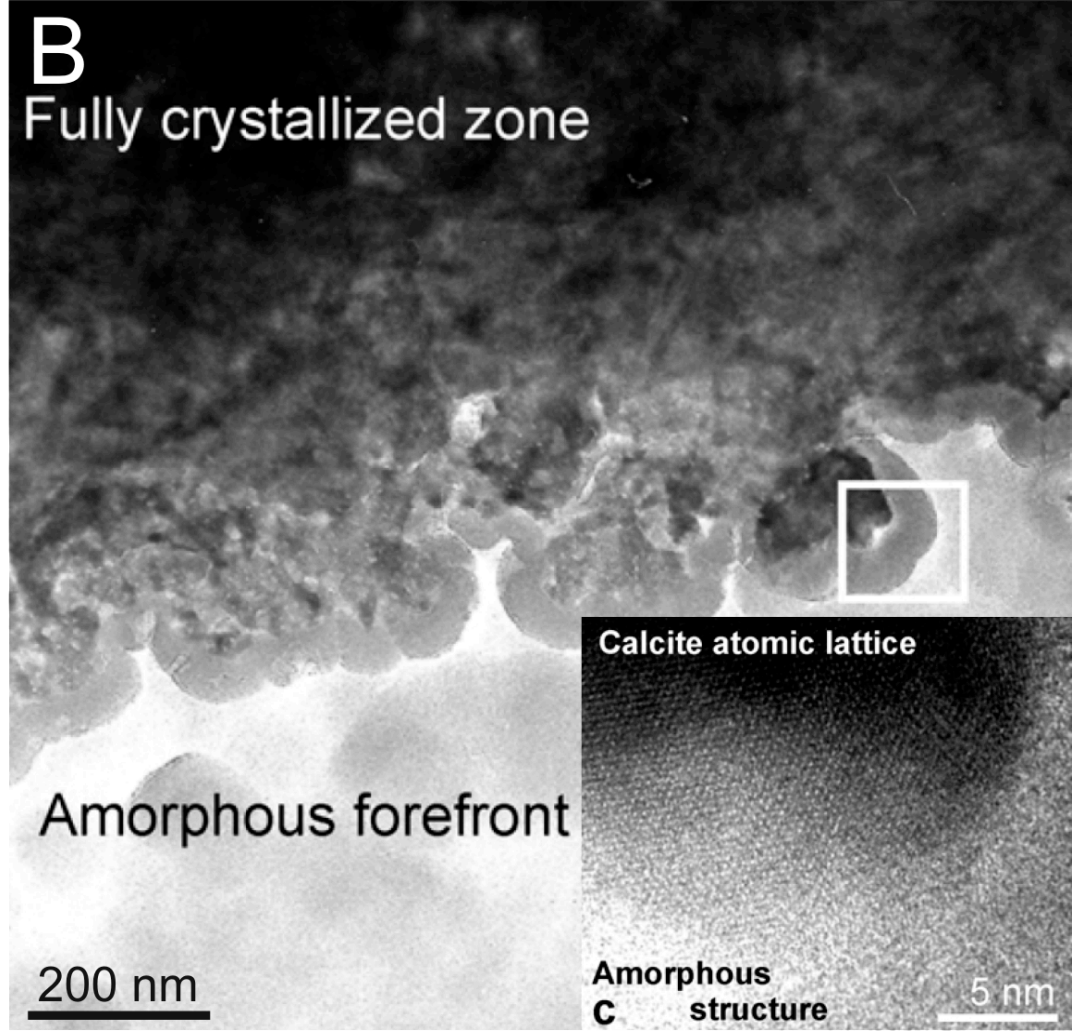

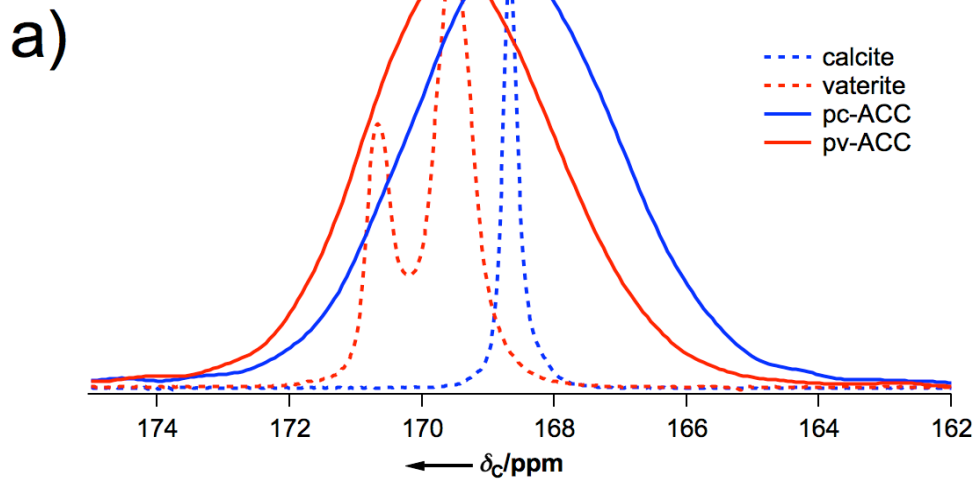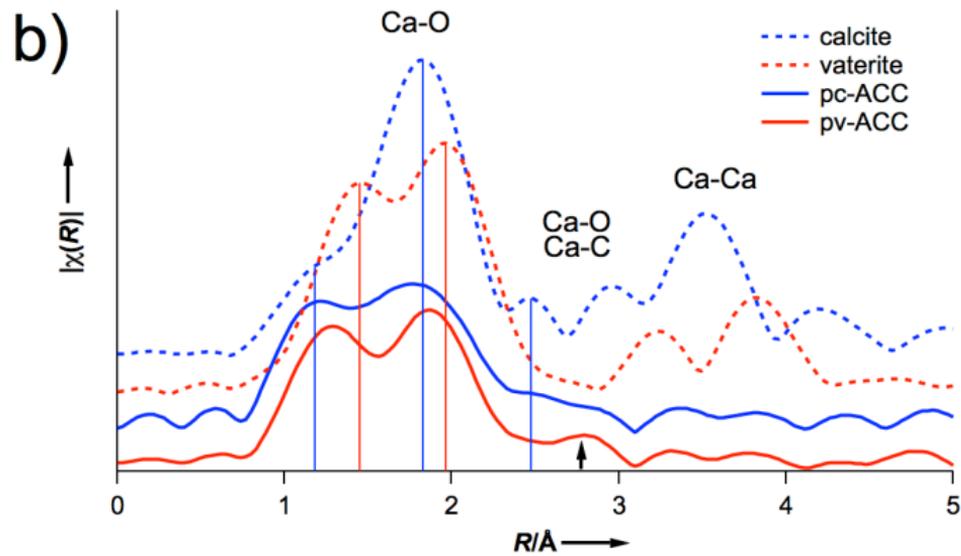

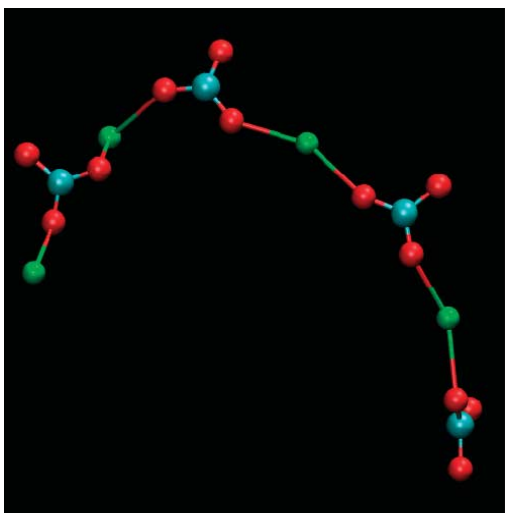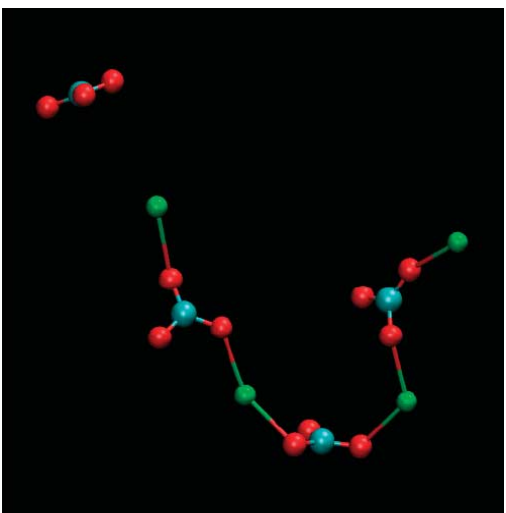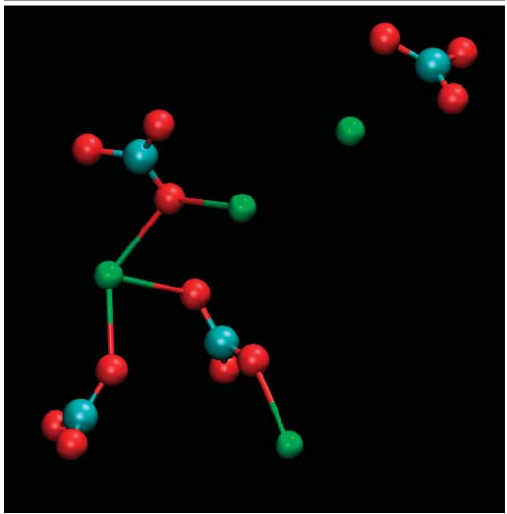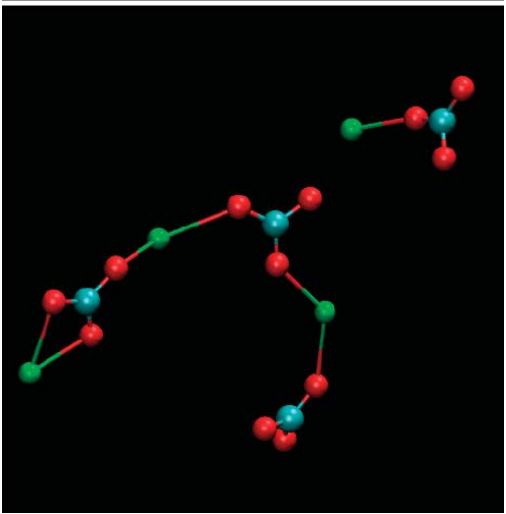

# CaCO₃ Polyamorphism

| biogenic |

- hydrous ACC
- calcitic ACC
- MHC-like ACC
- aragonitic ACC
- vateritic ACC(?)
- an-hydrous ACC

| stabilized |                                                        | transient |

- un-structured ACC
- liquid ACC
- PILP
- pc-ACC
- pv-ACC
- pa-ACC(?)

| synthetic |

## precursors to polymorphs?

| hydrous |

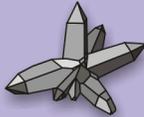

**Ikaite**
monoclinic

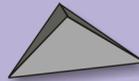

**Monohydrocalcite**
trigonal

| anhydrous |

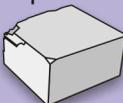

**Calcite**
trigonal

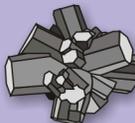

**Aragonite**
orthorhombic

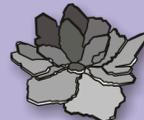

**Vaterite**
hexagonal (?)